\documentclass[10pt,letterpaper]{revtex4-1}
\usepackage{graphicx}
\usepackage{amsmath}
\usepackage{textcomp}
\usepackage{mathrsfs}

\begin{document}

\title{Analytical Approximations of Whispering Gallery Modes in Anisotropic Ellipsoidal Resonators}

\author{Marco Ornigotti$^1$}
\author{Andrea Aiello$^{2,3}$}
\address{$^1$Institute of Applied Physics, Friedrich-Schiller University, Jena, Max-Wien Platz 1, 07743 Jena, Germany} 
\address{$^2$Max Planck Institute for the Science of Light, G$\ddot{u}$nther-Scharowsky-Stra$\ss$e 1/Bau24, 91058 Erlangen, Germany} 
\address{$^3$Institute for Optics, Information and Photonics, University of Erlangen-N$\ddot{u}$rnberg, Staudtstra$\ss$e 7/B2, 91058 Erlangen, Germany}
\email{marco.ornigotti@uni-jena.de}

\date{\today}

\begin{abstract}
Numerical evolutions of whispering gallery modes of both isotropic and anisotropic spheroidal resonators are presented and used to build analytical approximations of these modes. Such approximations are carried out mainly to have the possibility to have a manageable analytic formulas for the eigenmodes and eigenfrequencies of anisotropic resonators. A qualitative analysis of ellipsoidal anisotropic modes in term of superposition of spherical modes is also presented.
\end{abstract}

%Insert PACS here
\pacs{}

%\maketitle
\section{I. Introduction}
Thanks to their extremely high quality-factor (up to $10^{10}$), whispering gallery mode (WGM) resonators are nowadays very promising devices for applications in optoelectronics  \cite{reviewWGM}, bio-sensing \cite{bio1,bio2}, nonlinear optics \cite{Josef} and other fields of applied physics. The interested reader on possible applications of WGM resonators is addressed to Ref. \cite{reviewWGM} and references therein. 
Although the first analysis of such devices dates back to 1939 in a paper by Richtmyer \cite{Richtmyer}, this topic remained unconsidered for many years until recent times, when the need of compact optical devices with great performances, together with a technological boost that gave the possibility to manufacture not only silica microspheres but WGM resonators of any material (such as crystalline resonators) and shape (spheroidal \cite{sfera}, toroidal \cite{tori} and so on), had contributed to make WGMs to experience a new renaissance.
In these years, the theory of WGMs in microspheres was well established both analytically and numerically and allowed precise calculation of eigenmodes, radiative losses, field distributions, et cetera \cite{reviewWGM}. Unfortunately, when approaching the study of anisotropic devices (more interesting from the point of view of the applications due to their very good performances), analytical solutions cannot be found for geometries different from an ideal sphere \cite{miopaper} or cylinder: in this case, it is more convenient to solve the problem with the help of numerical codes.

Sometimes, however, an analytical solution (even if not so accurate as the numerical solution) would be preferred either because it is easier to manage or because one has the possibility to directly catch the physics of the problem by directly looking at its form. But how to validate an analytical approximation of a solution when a real analytical solution does not exist? To answer to this question, we propose in this paper to use a numerical finite element solver (COMSOL Multiphysics \cite{comsol}) to firstly find an appropriate approximation of the problem in exam and then use that approximation to check the validity of an analytical model for the solution. 

Although this method is potentially applicable to a vast class of problems, in this paper we focus our attention to a specific problem: finding whispering gallery modes (WGMs) in isotropic and anisotropic dielectric ellipsoidal resonators. Even if there are some works about the approximations of the WGMs for isotropic non-spherical bodies that present analytical approximations in terms of spherical-like functions \cite{Gorodetsky,GeomWGM} or superposition of spheroidal modes \cite{spheroidal},  to the best of our knowledge no papers have been written on analytical solution for the anisotropic case. We then intend to fill this gap by finding with COMSOL a suitable approximation for WGMs in these resonators and then presenting a correspondent robust and accurate analytical solution.
The approximation we present is based on the observation that since WGMs are localized near the resonator surface, an effective approximation of an ellipsoidal resonator near its surface could be represented by a toroidal resonator of circular cross section, the radius of the torus being equal to the major axis of the resonator and the radius of the circular cross section being equal to the rim radius of the resonator at the considered surface.

Within this framework we calculate both the field distributions and the eigenvalues of the WGMs of both ellipsoidal resonator and toroidal resonators, showing how the WGMs of the torus well approximate the WGMs of the ellipsoidal resonator. This result gives us the possibility to substitute, under certain conditions, the complete set of spheroidal wave functions \cite{spheroidal} that characterize the ellipsoidal resonator with the complete set of spherical wave functions \cite{miopaper,wgm}. This permits us to manage a simple and analytical set of eigenmodes that can be easily used for theoretical predictions of nonlinear or quantum optical effects in these resonators.

This paper is organized as follows: in section II the model used to simulate with COMSOL such a resonator is presented, together with a brief recall of the weak-form expression for Maxwell's equations in a rotationally-symmetric resonator that is used as a model for the calculations. Section III is then devoted to present the results of the COMSOL simulations we made both for the isotropic and anisotropic case, with a direct comparison between the field distributions and the eigenvalues of both the ellipsoidal and toroidal resonators. Finally, in section IV conclusions are presented.
\section{II. The Model}
\subsection{A. Geometry of the resonator}

\begin{figure}[!t]
\centering\includegraphics[width=\textwidth]{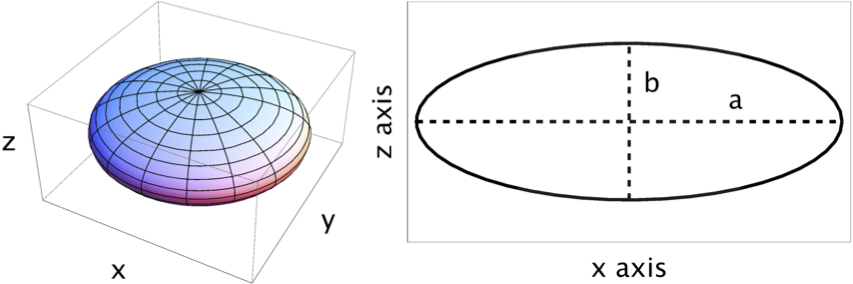}
\caption{(color online) Sketch of the 3D geometry of the resonator (left) and its cross section in the $xz$ plane (right). The numerical values of the semiaxes are $a=1.9$ mm and $b=0.7$ mm.The cross section on the $xy$ plane is a circle of radius $a$.}
\label{figura1}
\end{figure}

Let us consider an ellipsoid of revolution filled by a dielectric medium, rotationally invariant around the $z$-axis and characterized by the in-plane major axis $a$, i.e. the radius of the circumference in the plane $xy$, and the out-of-plane minor axis $b$, i.e. the minor axis of the ellipse in the plane $xz$. Such an ellipsoid is depicted in Fig.\ref{figura1}, and its cartesian equation is the following:\\
\begin{equation}\label{ellissoide}
\frac{x^2}{a^2}+\frac{y^2}{a^2}+\frac{z^2}{b^2}=1,
\end{equation}
where the major axis is $a=1.9$ mm and the rim radius $\rho=0.25$ mm; the value of the minor axis is then determined from these two numerical values with easy geometrical considerations and it turns out to be $b=0.7$ mm\cite{rim}. Following  Ref.\cite{Josef}, we consider this resonator to be made of Lithium Niobate ($\mathrm{LiNbO_3}$), an anisotropic crystal characterized by the following dielectric tensor:\\
\begin{eqnarray}\label{epsilon}
\hat{\varepsilon}& =& \left(
\begin{array}{ccc}
\varepsilon_{\perp} & 0 & 0\\
0 & \varepsilon_{\perp} & 0\\
0 & 0 & \varepsilon_{\parallel}\\
\end{array}
\right),
\end{eqnarray}
where $\varepsilon_{\perp}=5.3$ is the dielectric constant in the $xy$ plane and $\varepsilon_{\parallel}=6.5$ is the dielectric constant parallel to the $z$ axis.  However, in considering the resonator as isotropic, we assume that its dielectric tensor will be diagonal, with nonzero elements that are equal to $\varepsilon_{\perp}$.

\begin{figure}[!t]
\centering\includegraphics[width=\textwidth]{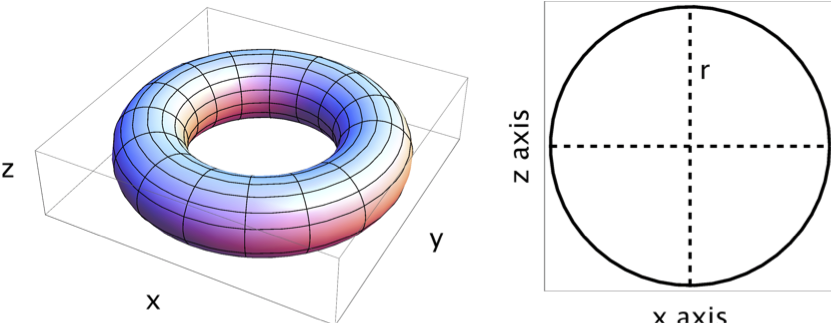}
\caption{(color online) Sketch of the 3D geometry of the toroidal (left) resonator that approximates the ellipsoidal resonator and its cross section in the $xz$ plane. The radius of the circular cross section is $r=\rho=0.25$ mm, and the external radius of the torus is equal to $a=1.9$ mm.}
\label{figura2}
\end{figure}

\begin{figure}[!t]
\centering\includegraphics[width=\textwidth]{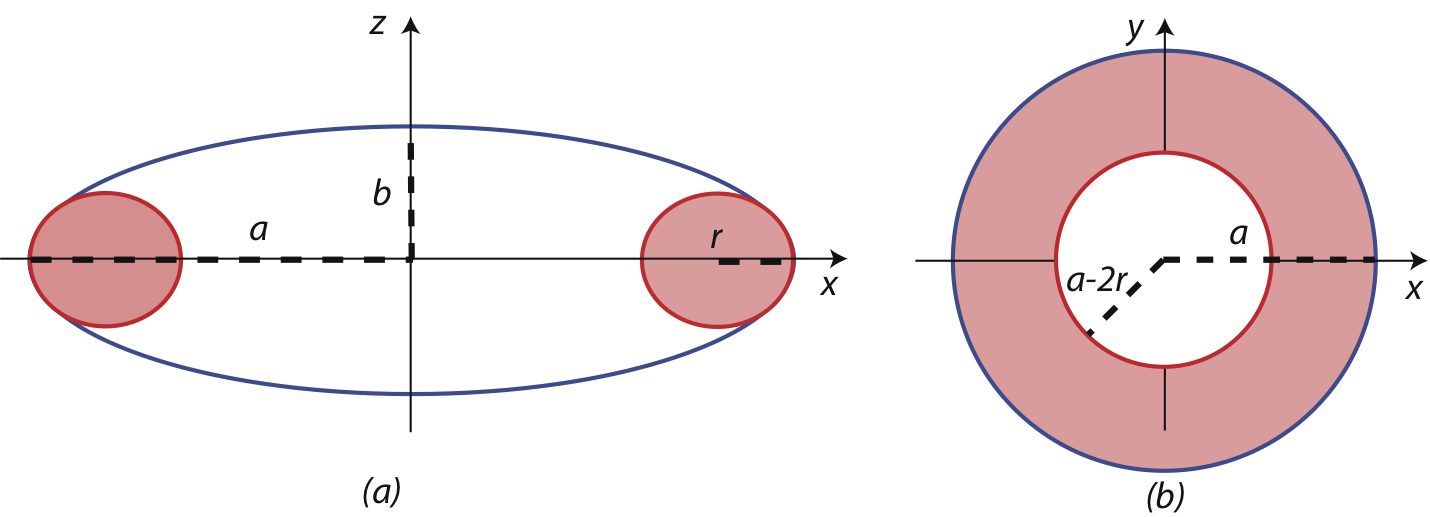}
\caption{(Color Online) Lateral (a) and top (b) view of the ellipsoidal resonator (blue solid line) and the approximating torus of circular cross section (red shaded circle). The radius of the torus cross section $r$ matches the rim radius $\rho=b^2/a$ of the ellipsoid.  The external radius of the torus matches the major axes of the ellipsoid (that appear as a circle of radius $a$ in the top view), while the internal radius of the torus is such that its circular cross section has a radius of $r$.}
\label{figuraApprox}
\end{figure}

In order to make a suitable approximation, we compare the results obtained from this resonator with the ones obtained from a toroidal resonator with circular section like the one depicted in Fig. \ref{figura2}. Here the external radius of the torus being $a=1.9$ mm equal to the major axis of the ellipsoidal resonator, and the radius of the circular cross section being $r=\rho=0.25$ mm, equal to the rim radius of the ellipsoidal resonator. A sketch of this approximation is shown in Fig. \ref{figuraApprox}, where both the lateral and top sections of the two resonators are shown.


  Even if an analytical solution for the open ellipsoid (with the fields continuous at the resonator surface) does not exist, it is possible to transform the open resonator into a closed one (with ideal metallic boundaries, i.e., the electric field is zero at the resonator surface) by simply considering that the field of a WGM outside the resonator is evanescent (i.e. exponentially decaying from the resonator surface) and replace the real spheroid with semiaxes $a$ and $b$ with effective semiaxes $\bar{a}=a+\sigma$ and $\bar{b}=b+\sigma$, where $\sigma$ is the depth upon which the optical field penetrates in the surrounding medium under total internal reflection. For grazing angles, $\sigma$ does not depend on the angle and has the value:\\
\begin{equation}
\sigma=\frac{k_0}{\chi\sqrt{\varepsilon-1}},
\end{equation}
where $k_0$ is the vacuum wavenumber, $\chi=1$ for quasi-TE modes and $\chi=\varepsilon$ for quasi-TM mode, being $\varepsilon$ the dielectric constant of the isotropic resonator. A detailed analysis on the origin of this term can be found in Ref. \cite{GeomWGM}. For the rest of the paper we will implicitly assume that the resonator is closed with semiaxes $a$ and $b$, remembering that the solutions we will present can be easily adapted to the open resonator by simply substituting $a$ and $b$ with $\bar{a}$ and $\bar{b}$.

\subsection{B. Weak formulation of the electromagnetic problem}

Before entering deep in the subject of this paper, we briefly intend to recall to the mind of the reader the weak formulation of Maxwell's equations with Galerkin's method of the weighted residuals, that we use to simulate with COMSOL whispering gallery modes in a general axisymmetric dielectric medium with permittivity tensor $\varepsilon$. 

We choose to implement such a model rather than using a standard mode solver  because the latter cannot be easily configured to fully exploit the axial symmetry of the problem and they experience some problems when dealing with WGMs. Then in order to obtain from them an accurate solution a fully 3D model must be implemented, making the calculations very time consuming and complicated. 

Galerkin's method, on the other hand, has the advantage that can easily take into account the symmetry of the problem, giving in this case the possibility to reduce the dimension of the problem from 3D to 2D, by solving the problem only in the $xz$ plane thanks to the axial rotation symmetry along the $z$ axis of the resonator. 

This method, compared with mode solvers, allow us to save a lot of computational time and memory, and gives us the possibility to calculate all the fields in the post processing phase of the simulation and only if we are interested in those quantities.

This method can be applied either for solving directly the electric field components \cite{weak1} or the magnetic field components \cite{weak2}, depending on what is the best choice with respect to the problem to be solved. 

Our results are based on the discussion presented in Ref. \cite{weak1}, where the calculation of the electromagnetic field distribution on an axisymmetric resonator is generalized to an arbitrary number of dielectric media embedded in an outer dielectric resonator, and they will be particularized in this work to study the simpler case of a resonator made by only one dielectric medium. 

The electromagnetic field inside the resonator obeys Maxwell's equations in continuous macroscopic media \cite{Jackson}. For this reason, in general, if one assumes that the resonator's constituent medium has constant magnetic susceptibility, then the magnetic field $\mathbf{H}$ will be continuous across the resonator volume, and the problem can be easily solved in terms of the magnetic field rather than the electric field $\mathbf{E}$. 

In the general case, however, the medium is not isotropic, and the dielectric constant became a tensor. For the sake of clarity, we will approach the problem from a general point of view, leaving the dielectric constant as a tensor in order to obtain a general expression for the magnetic field inside such a resonator. 

After obtaining the general expression, we will specialize it to either the case of isotropic (Section III.A) or anisotropic (Section III.B) resonator.

By eliminating the electric field in favor of the magnetic field in the Maxwell's equations, after some simple algebra the problem reduces to solve the following equation:\\
\begin{equation}\label{maxwell}
\nabla\times\Big(\hat{ \varepsilon}^{-1}\nabla\times\mathbf{H} \Big)-\alpha\nabla\Big(\nabla\cdot\mathbf{H}\Big)+\frac{1}{c^2}\frac{\partial^2}{\partial t^2}\mathbf{H}=0,
\end{equation}
where $\alpha$ is the so-called \emph{penalty} constant that controls the presence of spurious solutions (i.e. solutions with nonzero divergence inside the resonator) that may arise during the computation. The presence of this term is quite common when a weak form differential problem has to be solved with the help of finite element codes (see for instance Ref. \cite{penalty}). For all the simulations presented in this paper, the penalty constant is equal to one. $\hat{\varepsilon}^{-1}$ is the inverse relative permittivity tensor (sometimes named in literature as the  impermeability tensor), assumed to be independent of field strength. 

In order to have the complete solution of this equation, suitable boundary conditions must be applied: in our case we will assume that the resonator is closed and so we will use the so-called electric wall boundary conditions, namely:\\
\begin{subequations}
\begin{equation}
\mathbf{H}\cdot\mathbf{n}=0,
\end{equation}
\begin{equation}
\mathbf{E}\times\mathbf{n}=0.
\end{equation}
\end{subequations}
 These two equations imply that the magnetic field must be tangent to the resonator surface, while the electric field is normal to the resonator surface.
 
 From Eq. \eqref{maxwell}, and with the aid of Galerkin's method of weighted residuals \cite{Galerkin} it is possible to obtain the desired weak form of the electromagnetic problem, that will be implemented in COMSOL to be solved.
 
In order to reduce Eq. \eqref{maxwell} in its weak form counterpart, we multiply both sides of Eq. \eqref{maxwell} by the complex conjugate of a \emph{test} magnetic field $\bar{\mathbf{H}}^*$ and then integrate over the resonator's dielectric volume.

 Then we integrate by parts over the spatial coordinates and after the disposal of the surface integrals with the help of boundary conditions, we finally arrive to the desired weak form of Eq.\eqref{maxwell}, that reads:\\
  \begin{equation}
 \int_V\Big[\Big(\nabla\times\bar{\mathbf{H}}^*\Big)\cdot\Big(\frac{1}{\hat{\varepsilon}}\nabla\times\mathbf{H}\Big)
 -\alpha\Big(\nabla\cdot\bar{\mathbf{H}}^*\Big)\Big(\nabla\cdot\mathbf{H}^*\Big)\label{weakForm}
 \frac{1}{c^2}\bar{\mathbf{H}}^*\cdot\frac{\partial^2\mathbf{H}}{\partial t^2}\Big]dV=0,
 \end{equation}
 where the integral is taken over the whole resonator volume, and the first term under the integral stands for $\Sigma_{i,j=1}^3[\nabla\times\bar{\mathbf{H}}^*]_i\hat{\varepsilon}^{-1}_{ij}[\nabla\times\mathbf{H}]_j$. The three terms appearing in the integrand correspond exactly to the weak-form terms required to define an appropriate model within a partial differential equation solver \cite{Galerkin}. 
 
The axial symmetry of the problem suggests to describe the resonator in cylindrical coordinates $\{ r,z,\varphi \}$. With this choice of coordinate system the above equation turns into a 2D equation, because the $\varphi$-dependent part of the magnetic field $\mathbf{H}$ can be factored out in the form of a complex exponential $\exp{(iM\varphi)}$, where $M$ is the azimuthal quantum number, i.e.. the number of nodes of the WGM in the $xy$ plane. The total magnetic field can be then written as $\mathbf{H}(\mathbf{r})=\exp{(iM\varphi)}\Big[H_r(r,z)\hat{r}+H_z(r,z)\hat{z}+iH_{\varphi}(r,z)\hat{\varphi}\Big]$, where the imaginary unit in the $\varphi$-component is added to allow writing all the three components of the magnetic field in terms of an amplitude multiplied by a common phase factor. As it can be seen, the initially full 3D problem has been reduced to a 2D problem, thus resulting in a simplification of the problem and a speeding up of the computational time needed to execute the calculations.

The next step is to transform Eq. \eqref{weakForm} and its solution to a form easy to implement in COMSOL; this procedure is clearly presented in detail in Ref.\cite{weak1}, so we address the interested reader to this paper. We just point out that for the isotropic case one has to put $\varepsilon_{\perp}=\varepsilon_{\parallel}$ in Eq. (8) and following in Ref. \cite{weak1} ,  being $\varepsilon_{\perp}$ the dielectric constant in the $xy$ plane of the resonator and $\varepsilon_{\parallel}$ the dielectric constant parallel to the $z$-axis. In the case of uniaxial resonator, conversely, these two quantities have to be different.
 
\section{III. Results}
\subsection{A. Isotropic resonator}

\begin{figure}[!t]
\centering\includegraphics[width=\textwidth]{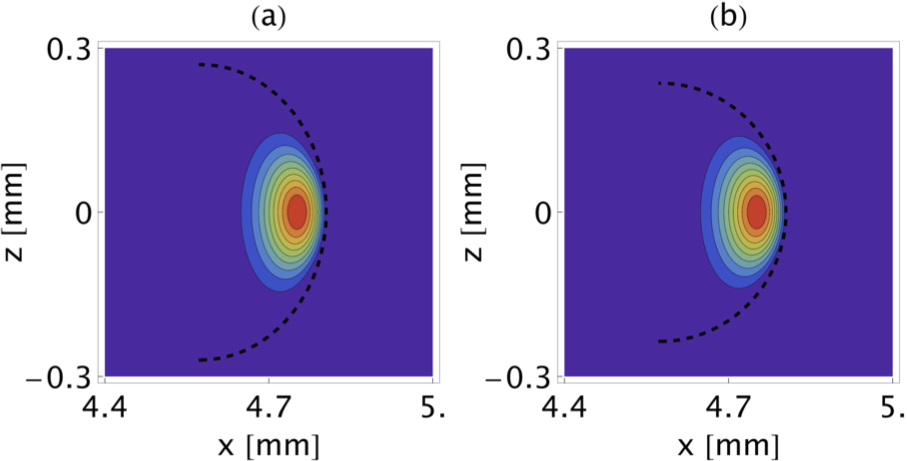}
\caption{(color online) Contour plot of the fundamental isotropic WGM ($m=1000$, $l-m=0$, $q=1$) of (a) the ellipsoidal resonator and (b) its toroidal approximation. The dashed black lines represent the resonators boundary.}
\label{modo1iso}
\end{figure}

\begin{figure}[!t]
\centering\includegraphics[width=\textwidth]{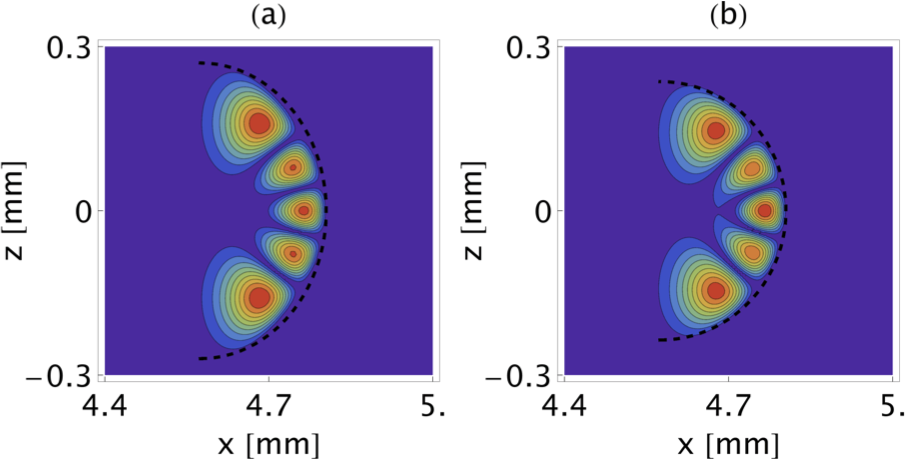}
\caption{(color online) Contour plot of the sixth isotropic WGM ($m=1000$, $l-m=4$, $q=2$) of (a) the ellipsoidal resonator and (b) its toroidal approximation. The dashed black lines represent the resonators boundary.}
\label{modo6iso}
\end{figure}

In Figs. \ref{modo1iso} and \ref{modo6iso} two examples of the field distribution for WGMs in the cross sectional $xz$ plane for both spheroidal (left) and toroidal (right) resonator are shown. In particular, Fig. \ref{modo1iso} shows the fundamental WGM characterized by the three quantum numbers $l=m=1000$ and $q=1$, being $l$, $m$ and $q$ the orbital, azimuthal and radial quantum numbers respectively. Fig \ref{modo6iso} instead shows the sixth WGM characterized by having $l-m=4$ and $q=2$. 

\begin{figure}[!t]
\centering\includegraphics[width=\textwidth]{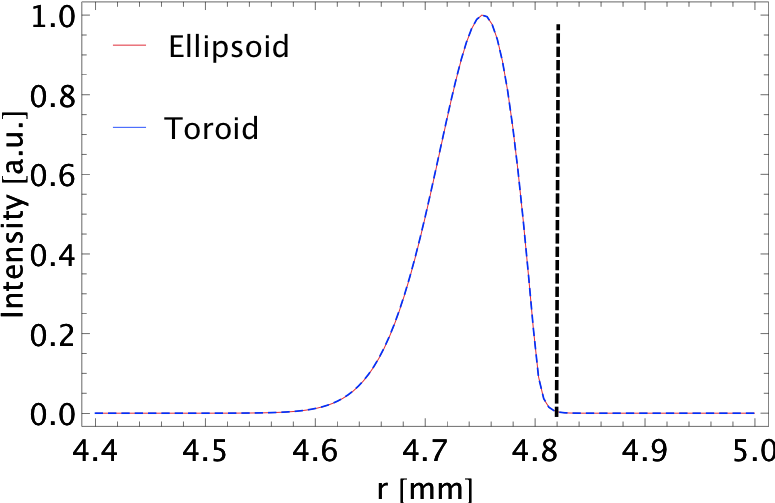}
\caption{(color online) Comparison of the x-section (radial part of the WGM) of both ellipsoidal (red curve) and toroidal (blue curve) fundamental WGM. The black dashed line represents the resonator boundary.}
\label{rSect1iso}
\end{figure}

\begin{figure}[!t]
\centering\includegraphics[width=\textwidth]{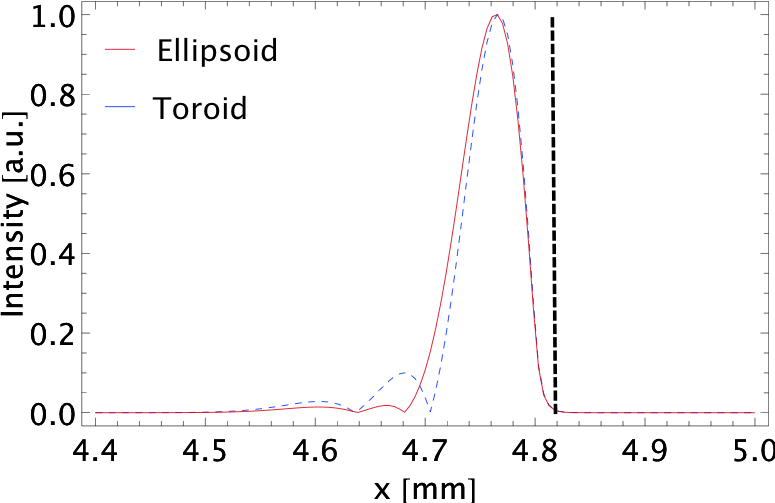}
\caption{(color online) Same as Fig. \ref{rSect1iso} but for the sixth WGM.}
\label{rSect6iso}
\end{figure}

\begin{figure}[!t]
\centering\includegraphics[width=\textwidth]{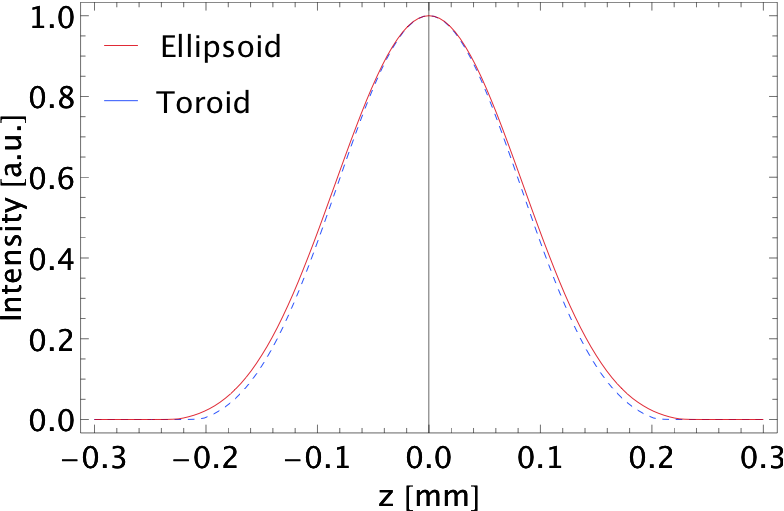}
\caption{(color online)Comparison of the z-section (angular part of the WGM) of both ellipsoidal (red curve) and toroidal (blue curve) fundamental WGM.}
\label{zSect1iso}
\end{figure}

\begin{figure}[!t]
\centering\includegraphics[width=\textwidth]{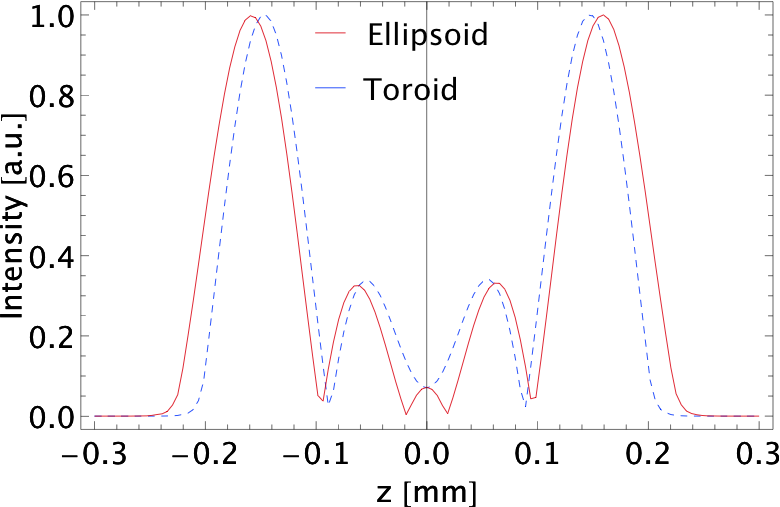}
\caption{(color online) Same as Fig. \ref{zSect1iso} but for the sixth isotropic WGM.}
\label{zSect6iso}
\end{figure}

As can be seen from these figures, and from their $x$- and $z$-sections shown in Figs. \ref{rSect1iso}-\ref{zSect6iso}, the toroidal approximates  the isotropic spheroidal resonator very well, resulting in the possibility to describe the set of spheroidal modes with the equivalent set of toroidal modes. Since here we are analyzing a 2D problem, and all that matters are the shapes of the modes in the cross section plane $xz$, saying that the toroidal modes well approximate the spheroidal mode is equal to say that these modes are very well approximated by the eigensolutions of the torus.


\begin{figure}[!t]
\centering\includegraphics[width=\textwidth]{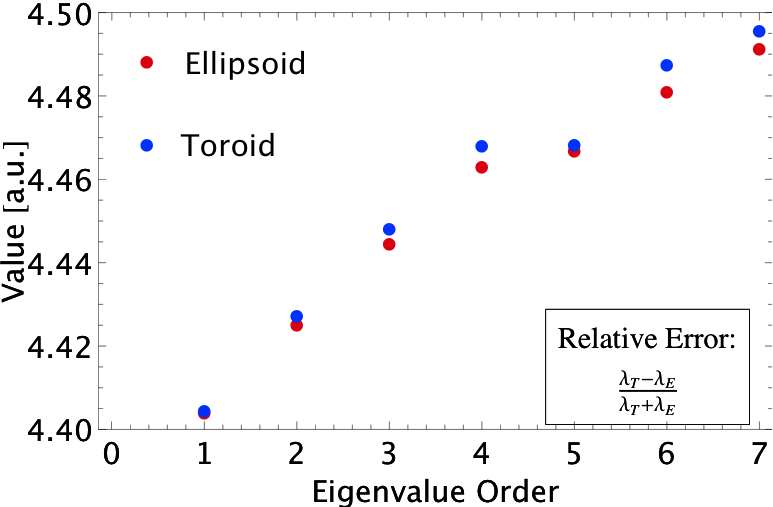}
\caption{(color online) Comparison between isotropic ellipsoidal (red dots) and toroidal (blue dots) eigenvalues corresponding to the first seven WGM. The inset shows the expression of the relative error used in this work to estimate the validity of our approximation. $\lambda_T$ is the eigenfrequency of the torus, while $\lambda_E$ is the eigenfrequency of the ellipsoidal resonator.}
\label{isoFreq}
\end{figure}

In order to make the approximation complete, we calculated the eigenvalues corresponding to the resonator's eigenmodes. Fig. \ref{isoFreq} shows a comparison between the eigenvalues of the first few WGMs of both spheroidal (red) and toroidal (blue) resonators. As can be seen, also in this case the approximation holds very well.
For the analysis and comparison of the eigenvalues of these two cavities, we adopted as a merit parameter for this approximation to hold the mean error on the estimation of the spheroidal eigenvalues with the toroidal ones (i.e. the numerical average of the error on every single eigenvalue). This mean error takes a value roughly around $0.3\%$ for the isotropic resonators, i.e. on the order of $O(l^{-1})$ for the WGMs presented here. 
Normally one can find in literature \cite{resonances} that the most common level of accuracy in determining the eigenvalues of a resonator is of $O(l^{-1})$ for WGMs. In our case this means that the precision needed for the eigenvalues must be of the order of $0.1\%$, that is precisely the accuracy lever of the toroidal approximation.

The approximation is very good for the first few WGMs, where the mode is highly confined near the resonator surface both in $x$ and $z$ direction. This results in a complete equivalence between the ellipsoidal and toroidal boundary condition. As can be seen from Fig. \ref{isoFreq}, the eigenfrequencies are very close to each other, resembling the fact that the modes have a very good overlap. This can be justified by also saying that while the modes of both resonators are highly confined near the resonator surface, they experience the same boundary conditions, i.e. they are almost identical.

When higher order modes are considered (like for example the one depicted in Fig. \ref{zSect6iso}), some discrepancies start to appear, since the mode starts to penetrate deep in the resonator. Formally, this approximation breaks when the considered mode is too extended in the resonator (namely along the $x$-direction) in such a way that its confining region is greater with respect to the region in which there is a good overlap between the curvature of the ellipse and the curvature of the circular cross section of the torus. In this case, the modes of the two cavities feel a different boundary (one feels ellipsoidal boundaries while the other one feels circular boundaries), and the discrepancy between these two modes starts to be significative.

\subsection{B. Anisotropic resonator}
\begin{figure}[!t]
\centering\includegraphics[width=\textwidth]{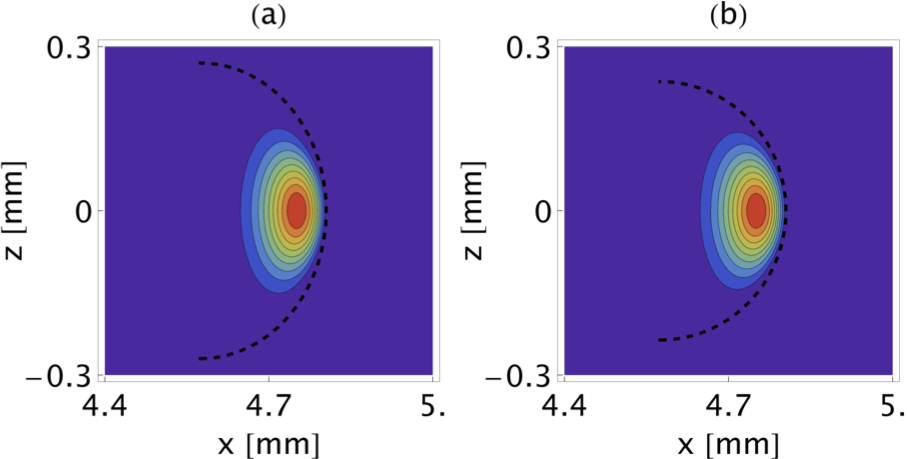}
\caption{(color online) Same as Fig.\ref{modo1iso} but for the anisotropic case.}
\label{modo1aniso}
\end{figure}

\begin{figure}[!t]
\centering\includegraphics[width=\textwidth]{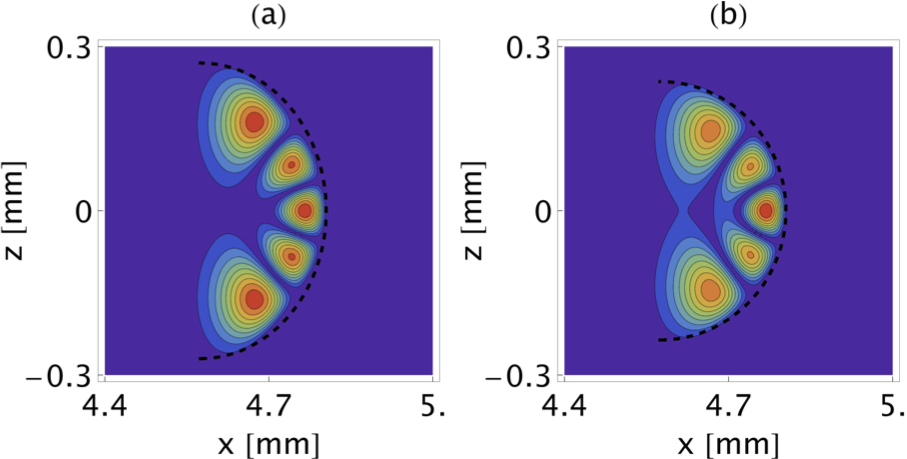}
\caption{(color online) Same as Fig. \ref{modo6iso}  but for the anisotropic case.}
\label{modo6aniso}
\end{figure}

\begin{figure}[!t]
\centering\includegraphics[width=\textwidth]{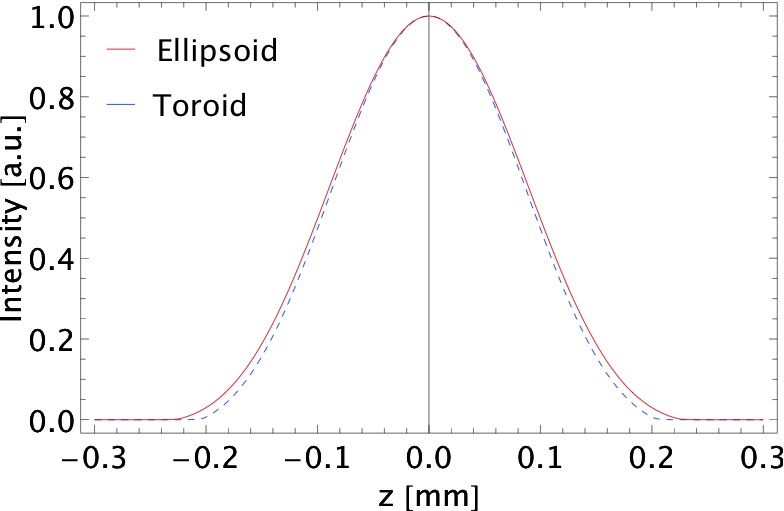}
\caption{(color online) Same as Fig. \ref{zSect1iso} but for the anisotropic case.}
\label{zSect1aniso}
\end{figure}

For the case of anisotropic resonators, results are reported in Fig.\ref{modo1aniso} and \ref{modo6aniso} for the bidimensional $xz$ cross sections of both spheroidal and toroidal modes and in Figs.\ref{rSect1aniso}-\ref{zSect6aniso} for their sections along the $x$- and $z$-axes respectively. 

The geometry of the anisotropic resonators is the same as the one described before for the isotropic case. The value of the dielectric function in the $xy$ plane and the one parallel to the $z$ direction are assumed to be $\varepsilon_{\perp}=5.3$ and $\varepsilon_{\parallel}=6.5$ respectively, as stated in section II.A.

\begin{figure}[!t]
\centering\includegraphics[width=\textwidth]{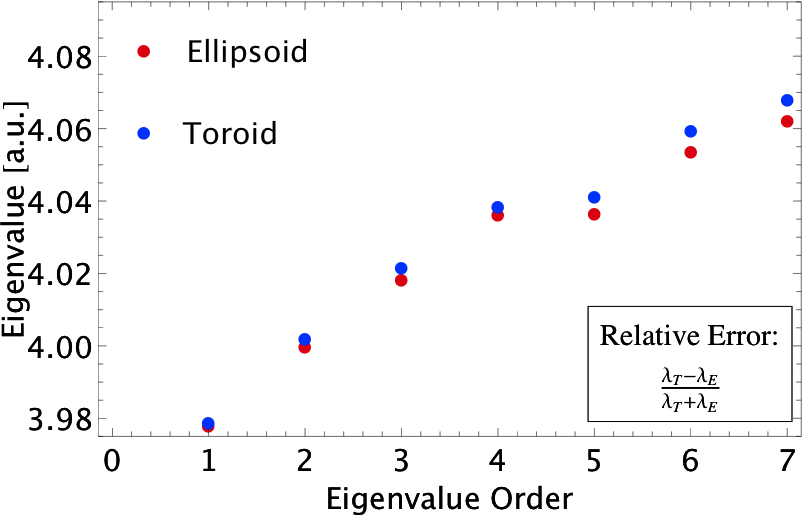}
\caption{(color online) Comparison between anisotropic ellipsoidal (red dots) and toroidal (blue dots) eigenvalues corresponding to the first seven WGMs. The inset shows the expression of the relative error used in this work to estimate the validity of our approximation. $\lambda_T$ is the eigenfrequency of the torus, while $\lambda_E$ is the eigenfrequency of the ellipsoidal resonator.}
\label{anisoFreq}
\end{figure}

Figure \ref{anisoFreq} shows instead the eigenvalue comparison for the anisotropic case for the first seven WGMs. As can be seen, even in the anisotropic case the toroidal approximation very well reproduces the eigenvalue pattern of the spheroidal resonator, and in this case the mean error was estimated to be roughly around $0.4\%$.

\subsection{C. Analytic formulas}
The approximations shown in the previous sections are of great interest, since they give us the possibility to approximate with a very good level of accuracy the WGM solutions of Helmholtz equations for an ellipsoid, by means of the WGM solutions of the Helmholtz equation in a toroidal cavity with circular cross section. This is very important from the theoretical point of view, because, as the WGM eigenmodes of the torus approximate well the WGM eigenmodes of the ellipsoid, then it is possible to use simple analytic expressions (as, for example, the solutions of the Helmholtz equation for the circle) to describe both isotropic and anisotropic ellipsoidal resonators.

The results presented here show that if we take a circumference whose radius is equal to the rim radius of the actual spheroid near its boundary, the solutions for this geometry very well approximate the $xz$ cross sections of the spheroidal solutions. In order to describe also the rotational symmetry of the spheroid, this circumference must be rotated around the $z$-axis, generating a toroidal resonator like the one depicted in Fig. \ref{figura2}. For this resonator, the $\varphi$-dependent term in the eigenmodes simply consists (due to the axial symmetry) in the complex exponential of the form $\exp{(iM\varphi)}$, where $M$ is the number of nodes of the eigenmode in the $xy$-plane, i.e. around the $z$-axis. This ansatz highly simplifies the analytical managing of such modes in these resonators, giving the possibility to work with the complete set of circular solutions (Bessel functions along the $x$-direction and trigonometric functions along the $z$-direction) rather than the more complicated spheroidal wavefunctions, concurring in a huge simplification of analytical calculations in such a system. Such calculations usually represent very hard tasks, as they require to manage infinite series of spherical solutions \cite{spheroidal}. It has to be pointed out, however, that this kind of approximation is not really needed when dealing with isotropic resonators, since a variety of analytical formulas is already available \cite{Gorodetsky, Abramowitz, GeomWGM}. The present approximation, on the other hand, is of great importance for anisotropic resonators, for which an analytical expression for the eigenmodes does not exist.

Having this goal in mind, let us first consider the radial wave equation for the circle, that has the following form:\\
\begin{equation}
\frac{d^2\psi}{dx^2}+\frac{1}{x}\frac{d\psi}{dx}+\Big(1-\frac{l^2}{x^2}\Big)\psi=0,
\end{equation}
where $x=kr$.  The solutions to this equation are the Bessel functions of the first kind $J_l(x)$, as they are the only ones that are finite at the origin. Nevertheless, as usual, it is possible to approximate the Bessel functions of large argument with the correspondent Airy functions $\mathrm{Ai}(x)$ \cite{wgm}. By exploiting these large-limit solutions, it is possible to build an analytical approximation to the ellipsoidal eigenmodes in the form of suitable combinations of Airy function as follows:\\
\begin{equation}\label{expansion}
\Psi(x)=\sum_{k=0}^Nc_k\mathrm{Ai}\Big[\gamma\Big(x-\beta+x_k\Big)\Big],
\end{equation}
where $x_k$ is the $k$-th zero of the Airy function and the quantities $\gamma$ and $\beta$ are two scaling and fitting parameters, respectively, that are used to take into account both the effects of anisotropy ($\gamma$ parameter) and other displacement effects ($\beta$ parameter). The weighting coefficients $c_k$ has to be chosen in such a way that the sum can be faithfully truncated to the order $N$. With this expansion, it is then possible to address the reals WGMs of the ellipsoidal resonator with a \emph{finite} number of solutions of the circle with large argument.

\begin{figure}[!t]
\centering\includegraphics[width=\textwidth]{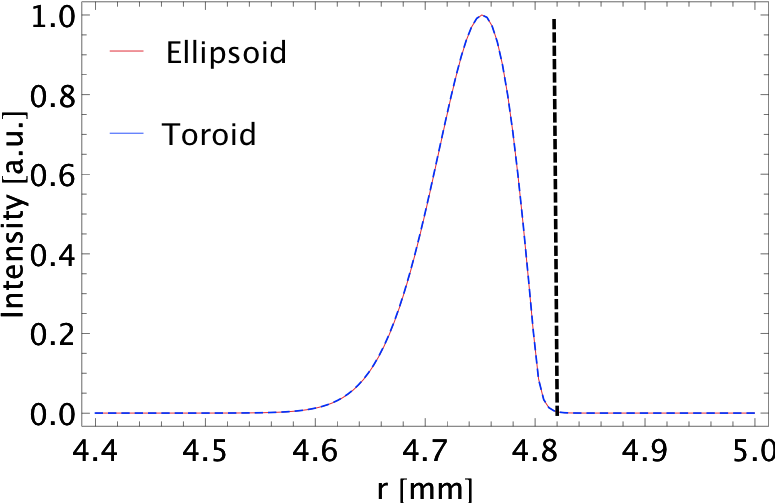}
\caption{(color online) Same as Fig. \ref{rSect1iso} but for the anisotropic case.}
\label{rSect1aniso}
\end{figure}

\begin{figure}[!t]
\centering\includegraphics[width=\textwidth]{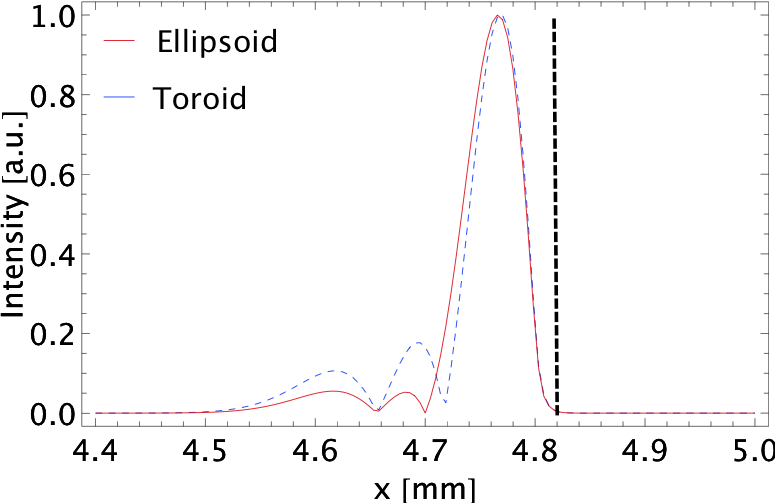}
\caption{(color online)Same as Fig. \ref{rSect6iso} but for the anisotropic case.}
\label{rSect6aniso}
\end{figure}

With these expressions for the WGMs it is then possible to fit the ellipsoidal modes. For example, the fundamental WGM described in Fig. \ref{rSect1aniso} can be very well reproduced by using only one term in the expansion \eqref{expansion} with $\gamma=25$ and $\beta=4.62$. In this case the term $x_k=-2.3381$ will be the first zero of the Airy function of the first kind $\mathrm{Ai}(x)$. For the more complicated pattern of Fig. \ref{rSect6aniso}, instead, the ellipsoidal WGM can be very well approximated by using three differently weighted Airy functions, so that $N=3$, with $\gamma=23$ and $\beta=4.46$. The weighting coefficient in this case are $c_1=0.15$, $c_2=-0.62$ and $c_3=1.8$.

Proceeding along this way, we have then the possibility to approximate, with a large degree of precision, all the WGMs of the ellipsoidal resonator (either isotropic or anisotropic) with a \emph{finite} number of eigensolutions of the Helmholtz equation in the circle, rather than using infinite series. 

This is of great advantage because it guarantees more manageable expressions, and allows one to easily understand the physics that lives behind these modes, as one can interpret them as just a linear combination of few modes from a circle.

In the same manner it is possible to analytically approximate the angular distribution of such modes, i.e.,  the WGM structure along the $z$ direction. For this case we have the possibility to employ  the solution to the angular equation for the circle\\
\begin{equation}
\frac{d^2\Theta}{dz^2}+l^2\Theta=0,
\end{equation}
 by writing again the angular part of the mode of the ellipsoid  as a finite sum of trigonometric terms:\\
\begin{equation}\label{angular_expansion}
\Theta(z)=\sum_{k=1}^N\Big[b_k\cos(\beta k z)+d_k\sin(\beta k z)\Big],
\end{equation}
where $\beta=l \alpha$ is the fitting parameter ($l$ being the orbital quantum number that appearing in the angular equation) that, again, takes into account a possible anisotropy even along the angular distribution, and $b_k$ and $d_k$ are weighting coefficients.

 It has to be pointed out, however, that while for an uniaxial anisotropy in a sphere the anisotropy will affect (at the first order) just the radial part of the mode \cite{miopaper}, in the case of an ellipsoid we must account for an anisotropy fitting parameter even in the angular part of the mode since, as the mode index grows, the mode starts to extend itself into the resonator, breaking the toroidal approximation, then starting to feel ellipsoidal boundary conditions rather than the circular ones. This can be considered and modeled as an effective anisotropy that acts prevalently along the angular direction.

\begin{figure}[!t]
\centering\includegraphics[width=\textwidth]{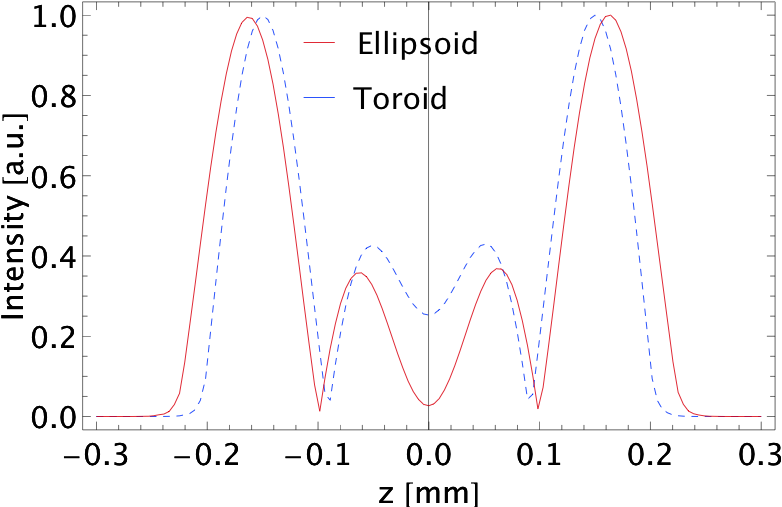}
\caption{(color online) Same as Fig. \ref{zSect6iso} but for the anisotropic case.}
\label{zSect6aniso}
\end{figure}

For example, for the sixth mode of the anisotropic ellipsoidal resonator as the one shown in Fig. \ref{zSect6aniso}, for which $l=1000$ and $l-m=4$, a very good analytical approximation can be built with the help of Eq. \eqref{angular_expansion} by considering only the even part of the expansion (i.e.,  only cosine terms), because the mode shown in Fig. \ref{zSect6aniso} possesses even symmetry with respect to the vertical axis. We then have $N=4$ and the expansion coefficients are $b_1=2.6$, $b_2=-1.5$, $b_3=0.1$, $b_4=-0.2$; the effective anisotropy parameter takes the value $\alpha=0.022$ and $d_k=0$ since we consider only angular function with even parity.

As can be seen, the value of the effective anisotropy parameter in the argument of the angular functions is very low but still different from zero. This means that while for a spherical anisotropic resonator (if the anisotropy is small) the angular function is not affected by the anisotropy, in the case of the anisotropic ellipsoidal resonator, this is also true for the low order WGMs. As the order of the WGM grows, than an effective anisotropy (arising from the different shapes of the spherical and ellipsoidal boundaries that the mode sees), starts to play an important role even in the angular distribution of the mode.
\section{IV. Conclusions}
In this work we have used a numerical finite element solver like COMSOL to find the field distribution of WGMs in a spheroidal resonator. We have proposed to approximate the spheroidal resonator near its boundaries with a toroidal resonator with circular cross section, having the radius equal to the major axis of the spheroid and the radius of the circular cross section equal to the rim radius of the spheroid neat its surface.  We presented results for both isotropic and anisotropic resonators and we pointed out that, especially for anisotropic resonators, thanks to the numerical verification of our approximation, it is possible to approximate WGMs of a spheroid with a suitable superposition of modes of the circle. The advantage of our proposed analytical approximation is two-fold: from one side it gives the possibility to easily represent the ellipsoidal WGMs as superposition of few circular eigenmodes rather than infinite series of spherical harmonics. On the other side, for the case of anisotropic resonator, it gives easy manageable formulas to describe modes that otherwise will not have an analytical representation.  This could be of great help when challenging complicated calculations on such resonators because at first it allows us to use analytical (i.e.,  simple and practical) methods to have a first educated guess about how things should go inside the considered system. Second, a posteriori, availability of simple approximate formulas, permits very powerful qualitative analysis of the quantitative numerical results.

% Create the reference section using BibTeX:

%
%
%

\begin{thebibliography}{99}
\bibitem{rim}
It can be shown with simple geometrical arguments that the rim radius of an ellipse at $\theta=0$ ($\theta$ being the angle formed by the position vector (giving the position of a point on the surface of the ellipse with respect of the center of the ellipse) and its major axis) is equal to $\rho=b^2/a$. This result is here used to infer from the data presented in Ref. \cite{Josef} the correct value of the minor axis $b$.
\bibitem{commento2}
For the case of the spheroid, for example, analytical solutions for the uniaxial spheroid are not available because the matrix elements of the dielectric tensor are functions of all the three spheroidal coordinates $\xi$, $\eta$ and $\varphi$ and a procedure similar to the one presented in Ref.\cite{miopaper} cannot be applied.
\end{thebibliography}
\end{document}